\def\arxivversion{1}
\documentclass[aps,prl,twocolumn,superscriptaddress,groupedaddress,nofootinbib,longbibliography]{revtex4-2}

\usepackage{graphicx}
\usepackage{bm}
\usepackage{amsmath}
\usepackage{amssymb}
\usepackage{hyperref}

\providecommand{\arxivversion}{0}
\newif\ifarxiv
\ifnum\arxivversion=1 \arxivtrue\fi

\begin{document}

\title{Electromagnetic Backgrounds and Angular-Momentum Barriers in Gravitomagnetic Searches with Hydrogen-Like Ions}

\author{Leonardo A. Pach\'{o}n}
\affiliation{guane Enterprises, R+D+I Unit, Medell\'in 050010, Colombia}

\date{October 18, 2026}

\begin{abstract}
Hydrogen-like ions cannot usefully bound the gyrogravitational ratio $\chi$: in ${}^{95}$Mo$^{41+}$ gravity shifts the hyperfine constants by fractions of $10^{-39}$--$10^{-38}$ with the tensor structure of the hyperfine interaction, centroids cancel it, and the $2p$ lines are 1.3 eV wide. Isotopes with spin can, however, enter the King plots that search for new bosons: for the cadmium clock line, the second-order hyperfine shift of ${}^{111}$Cd is $-0.96$~MHz, and a specific difference with ${}^{113}$Cd reduces it to about $2$~Hz after calculable anomaly and third-order corrections.
\end{abstract}

\maketitle

The gravitomagnetic field of a rotating mass has been measured for the Earth: Gravity Probe B measured the frame-dragging precession of orbiting gyroscopes to $19\%$~\cite{Everitt2011}, and the LAGEOS satellites the Lense--Thirring drag of their orbits~\cite{Ciufolini2004}. Its coupling to the intrinsic spin $\mathbf{S}$ of a test particle, $H=\chi\,\mathbf{S}\cdot\bm{\Omega}_{\rm LT}$, has $\chi=1$ in general relativity (GR)~\cite{Mashhoon2003,Obukhov2009}; we call $\chi$ the gyrogravitational ratio, and $\chi-1$ is the anomalous gravitomagnetic moment of Ref.~\cite{SilenkoTeryaev2007} (in the spin-sensor literature the same name is given to the coefficient of a coupling $\chi\,\bm{\sigma}\cdot\mathbf{g}$ of spin to the gravitational acceleration~\cite{Jiang2025}). The coupling of intrinsic spin to the field of a rotating source has not been measured~\cite{Fadeev2021}; it has been tested through the equivalent coupling of spin to the rotation of the laboratory. For nuclear spins, the comparison of ${}^{199}$Hg and ${}^{201}$Hg in the rotating Earth~\cite{Venema1992} bounds $|(\chi_{201}-1)+0.369\,(\chi_{199}-1)|<0.042$ at $95\%$ confidence~\cite{SilenkoTeryaev2007}, and a ${}^{129}$Xe/${}^{131}$Xe comagnetometer has measured the Earth's rotation rate to $3\times10^{-4}$~\cite{Zhang2023}; read as a test of the rotation coupling, this measurement bounds its fractional anomaly for the ${}^{129}$Xe/${}^{131}$Xe pair about 70 times more tightly than the Hg data~\cite{supplemental}. For electron spins, the gyrocompass torque of the Earth's rotation on a spin-polarized torsion pendulum confirms the coupling to $18\%$ ($1\sigma$)~\cite{Heckel2008,Ni2010}. Hydrogen-like ions such as ${}^{95,97}$Mo$^{41+}$ might seem to offer a test at the nucleus itself: their bound electron probes the nucleus at short distances, a gravitomagnetic coupling to the nuclear spin would act in every level, and a generalized King plot~\cite{Berengut2020} might separate it from electromagnetic effects.

Here we show that this program cannot work. The gravitational couplings of a nuclear spin to a bound electron are hidden behind four barriers, two set by angular momentum (I, III), one by the hyperfine interaction (II), and one by radiative decay (IV): (I)~quadrupole couplings require levels with $j\ge3/2$; (II)~within a level, every gravitational coupling has the tensor structure of a hyperfine interaction and only shifts the hyperfine constants; (III)~the hyperfine centroids used in isotope-shift analyses cancel it exactly; and (IV)~the $2p$ lines of hydrogen-like ions are too broad to resolve the hyperfine structure. The Supplemental Material (SM)~\cite{supplemental} gives the calculations.

\textit{Gravitational couplings.}---For any stationary source of mass $M$, GR makes the angular momentum $J=\hbar I$ the current (gravitomagnetic) dipole of the exterior field. For the Kerr metric, whose multipole moments are $M_\ell+iJ_\ell=M(ia)^\ell$ with $a=J/(Mc)$~\cite{Hansen1974}, $J$ also fixes the mass (gravitoelectric) quadrupole, $M_2=-J^2/(Mc^2)$, and there is no current quadrupole. At first order in $J$ the electron feels the Lense--Thirring field: its spin couples through $\mathbf{S}\cdot\bm{\Omega}_{\rm LT}$, with $\bm{\Omega}_{\rm LT}=G[3(\mathbf{J}\cdot\hat{\mathbf{r}})\hat{\mathbf{r}}-\mathbf{J}]/(c^2r^3)$ outside the origin~\cite{Mashhoon2003} and a contact term at $r=0$~\cite{supplemental}, and its orbital angular momentum $\mathbf{L}$ through $2G\,\mathbf{J}\cdot\mathbf{L}/(c^2r^3)$, which follows from the gravitomagnetic vector potential~\cite{Mashhoon2003,supplemental}. At second order it feels the Newtonian potential of the quadrupole. A static coupling of the electron spin to an $I^2$-dependent field is odd under time reversal and has no diagonal matrix elements~\cite{supplemental}.

What matters is the structure of these couplings. The Lense--Thirring coupling has the form of the magnetic-dipole (M1) hyperfine interaction, with orbital and spin parts weighted $2{:}1$ instead of $1{:}1$, and the quadrupole potential has the form of the electric-quadrupole (E2) one; outside the nucleus both fall off as $r^{-3}$. Within a level they therefore add to the hyperfine constants $A$ and $B$. For the $2p_{3/2}$ level of ${}^{95}$Mo$^{41+}$ ($I=5/2$), with the Dirac--Coulomb value of $\langle r^{-3}\rangle$ and nonrelativistic angular factors, accurate to $(Z\alpha)^2\approx0.09$~\cite{supplemental}, the Lense--Thirring coupling adds $A_{\rm grav}=1.4\times10^{-42}$~eV to $A=-2.43\times10^{-4}$~eV, and the spin-induced (Kerr) mass quadrupole adds $-2.6\times10^{-47}$~eV to $B=-2.78\times10^{-5}$~eV. A real nucleus is not a Kerr source: its mass quadrupole follows its deformation, which adds $B_{\rm grav}=-2.7\times10^{-44}$~eV if mass and charge have the same shape (for an odd-neutron nucleus such as ${}^{95}$Mo, only to within a factor of order unity). The ratios
\begin{equation}
\begin{split}
    \epsilon_1&=|A_{\rm grav}/A|=5.6\times10^{-39},\\
    \epsilon_2&=\frac{B_{\rm grav}}{B}=\frac{Gm_eM}{Ze^2/(4\pi\varepsilon_0)}=9.9\times10^{-40},
\end{split}
    \label{eq:ratios}
\end{equation}
change from level to level at most by factors of order unity and depend on the nucleus only through $I/\mu$ and $M/Z$ ($\mu=-0.9097\,\mu_N$ for ${}^{95}$Mo~\cite{IAEALiveChart}).

\textit{Barrier I: angular momentum.}---A coupling of nuclear rank $k$ has nonzero diagonal matrix elements in a level of angular momentum $j$ only if $k\le2j$~\cite{Varshalovich1988}. Quadrupole couplings, electromagnetic or gravitational, therefore require $j\ge3/2$; in a hydrogen-like ion the lowest such level is $2p_{3/2}$, and the $j=1/2$ levels see only the scalar and dipole parts.

\begin{figure}[t]
\includegraphics[width=\columnwidth]{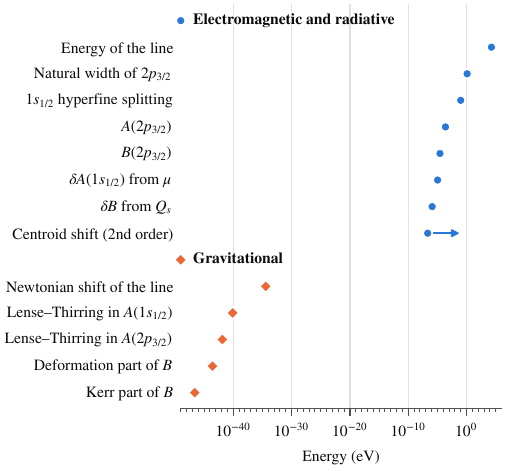}
\caption{\label{fig:scales} Energy scales of the $1s_{1/2}\to2p_{3/2}$ line of ${}^{95}$Mo$^{41+}$ (magnitudes, logarithmic axis). Circles: electromagnetic and radiative scales; diamonds: gravitational ones, with $\chi=1$. The centroid shift is a lower estimate (arrow), placed at the geometric mean of the second-order estimates of the SM~\cite{supplemental}. The other values are those of Table~\ref{tab:scales}.}
\end{figure}

\textit{Barrier II: degeneracy with the hyperfine interaction.}---Within a level of given $j$ and $I$, the Wigner--Eckart theorem fixes how any coupling of rank $k$ depends on the total angular momentum $F$: it follows the interval rule of the M1 constant $A$ for $k=1$ and the pattern of the E2 constant $B$ for $k=2$~\cite{supplemental}. A gravitational coupling therefore only shifts $A$ and $B$, by the fractions $\epsilon_1$ and $\epsilon_2$ of Eq.~(\ref{eq:ratios}). For the quadrupole, $\epsilon_2$ is the same in every level up to relative corrections of order $(Z\alpha)^2$, and to that accuracy the gravitational term is a renormalization of the nuclear quadrupole moment~\cite{supplemental}: for ${}^{95}$Mo$^{41+}$, the uncertainty of the spectroscopic quadrupole moment $Q_s=-0.022(1)$~b~\cite{StoneQ2021} alone makes $B(2p_{3/2})$ uncertain by $1.3\times10^{-6}$~eV, $5\times10^{37}$ times its gravitational part. For the Lense--Thirring term, whose orbital and spin parts are weighted $2{:}1$, $\epsilon_1$ differs from level to level by factors of order unity, so it acts as a hyperfine anomaly of relative size $\sim10^{-39}$~\cite{supplemental}, far below the Bohr--Weisskopf and QED corrections to the hyperfine constants. No choice of transition or charge state improves these ratios. Comparing isotopes does not help either: the gravitational quadrupole of a real nucleus is set by the same deformation as $Q_s$ and is not known independently of it, and the Kerr term is smaller still.

\begin{table}[t]
\caption{\label{tab:scales} Energy scales for the $1s_{1/2}\to2p_{3/2}$ line of ${}^{95}$Mo$^{41+}$, in eV (magnitudes; SM~\cite{supplemental}).}
\begin{ruledtabular}
\begin{tabular}{lc}
$1s_{1/2}\to2p_{3/2}$ transition energy & $1.8537\times10^{4}$ \\
Natural width of $2p_{3/2}$ & $1.25$ \\
Hyperfine splitting of $1s_{1/2}$ & $9.9\times10^{-2}$ \\
M1 hyperfine constant $A(2p_{3/2})$ & $2.43\times10^{-4}$ \\
E2 hyperfine constant $B(2p_{3/2})$ & $2.78\times10^{-5}$ \\
Uncertainty of $A(1s_{1/2})$ from $\mu$ & $1.1\times10^{-5}$ \\
Uncertainty of $B$ from $Q_s$ & $1.3\times10^{-6}$ \\
Second-order shift of the line centroid & $\gtrsim10^{-7}$ \\
Newtonian shift of the line, $GMm_e\Delta\langle r^{-1}\rangle$ & $3.8\times10^{-35}$ \\
Lense--Thirring part of $A(1s_{1/2})$, per unit $\chi$ & $8.0\times10^{-41}$ \\
Lense--Thirring part of $A(2p_{3/2})$ & $1.4\times10^{-42}$ \\
Deformation part of $B$ & $2.7\times10^{-44}$ \\
Spin-induced (Kerr) part of $B$ & $2.6\times10^{-47}$ \\
\end{tabular}
\end{ruledtabular}
\end{table}

\textit{Barrier III: centroids.}---Precision King plots use spinless isotopes to avoid hyperfine effects~\cite{Berengut2018,Counts2020,Hur2022}; when isotopes with nuclear spin are included, their hyperfine multiplets enter through centroids~\cite{BerengutOreshkina2025}. The $(2F+1)$-weighted centroid of a level is the trace of the interaction over its $(2j+1)(2I+1)$ states, and for every $k\ge1$
\begin{equation}
    \sum_F(2F+1)\,\langle(jI)F|\,\mathbf{t}^{(k)}\cdot\mathbf{T}^{(k)}|(jI)F\rangle=0,
    \label{eq:centroid}
\end{equation}
because the components of an irreducible tensor of rank $k\ge1$ are traceless; here $\mathbf{t}^{(k)}$ and $\mathbf{T}^{(k)}$ are rank-$k$ tensor operators acting on the electron and on the nucleus. The same holds for the centroid of an E1 line weighted by its line strengths~\cite{supplemental}. Centroids thus remove the M1 and E2 hyperfine structure at first order, which is why isotope-shift analyses use them, but they remove every gravitational coupling of rank 1 or 2 just as exactly. At first order, only gravitational terms independent of the nuclear spin survive in them. The largest is the Newtonian energy, which shifts the line by $3.8\times10^{-35}$~eV (Table~\ref{tab:scales}); this shift differs between ${}^{95}$Mo and ${}^{97}$Mo by $8.0\times10^{-37}$~eV, and the post-Newtonian and geodetic spin--orbit terms are smaller by factors of order $(Z\alpha)^2$~\cite{supplemental}. What the hyperfine interaction leaves in the centroid is of second order, and mostly M1--M1. In ${}^{95}$Mo$^{41+}$, mixing with $2s$ alone lowers the $1s$ centroid by $1.7\times10^{-8}$~eV; mixing with higher $ns$ levels and with the continuum, which dominates, lowers it further, to $\sim10^{-7}$~eV or more~\cite{supplemental}. Mixing of $2p_{1/2}$ with $2p_{3/2}$, $149.73$~eV apart, raises the $2p_{3/2}$ centroid by $3.6\times10^{-10}$~eV. In the intensity-weighted centroid of an unresolved line, the same mixing also moves intensity between components, which shifts it by $\gtrsim10^{-7}$~eV; the frequency dependence of the intensities shifts it by $\sim10^{-7}$~eV, and non-statistical populations of the hyperfine sublevels already at first order, by up to $4\times10^{-2}$~eV~\cite{supplemental}. Such second-order shifts are a known source of King-plot nonlinearity for isotopes with spin~\cite{BerengutOreshkina2025}. In a generalized King plot, odd-isotope residuals linear in $Q_s$, in a tensor nuclear polarizability or in $I^2/M$ would therefore have no first-order origin, since centroids cancel every rank-2 term at first order.

\textit{Barrier IV: line widths.}---The $1s_{1/2}\to2p_{3/2}$ line of Mo$^{41+}$ lies at $18.537$~keV, and its upper level decays in $0.53$~fs, a natural width of $1.25$~eV~\cite{NISTASD}; our Dirac calculation agrees within $5\%$~\cite{supplemental}. The hyperfine splittings, $0.099$~eV for $1s$ and at most a few $10^{-3}$~eV for $2p$, are unresolved, so these lines determine neither the $2p$ hyperfine constants nor the multiplet centroids at the level a gravitational test would require; the $1s$ splitting is accessible only through its own M1 transition. Quantum-logic spectroscopy~\cite{Schmidt2005,Micke2020} and long Ramsey times apply to narrow transitions between long-lived levels, such as the fine-structure transitions of few-electron ions~\cite{Kozlov2018}. Such levels with $j\ge3/2$ exist, but Barrier II applies to them unchanged.

\textit{Consequences for $\chi$.}---The gyrogravitational ratio of the electron multiplies only the spin part of the Lense--Thirring coupling. In the $1s$ level, whose splitting is measurable through its M1 transition, that part is the whole gravitational contribution, $A_{\rm grav}(1s)=8.0\times10^{-41}\chi$~eV; in the $2p_{3/2}$ level it contributes $-1.5\times10^{-43}$~eV to $A$, against $+1.5\times10^{-42}$~eV from the orbital part~\cite{supplemental}. A measured constant constrains $\chi$ only if its magnetic part can be predicted. For $1s$, the tabulated $\delta\mu/\mu=3.3\times10^{-4}$ of ${}^{95}$Mo~\cite{IAEALiveChart} alone makes that prediction uncertain by $1.1\times10^{-5}$~eV, so that at best $|\chi-1|\lesssim1.4\times10^{35}$; ratios of constants of different levels avoid $\mu$, but they would require the magnetic parts, including the Bohr--Weisskopf and QED corrections, to be predicted to $10^{-39}$ of $A$. This best case is $8\times10^{35}$ times weaker than the $18\%$ ($1\sigma$) test of the coupling of electron spins to the Earth's rotation~\cite{Heckel2008,Ni2010}. No meaningful laboratory bound on $\chi$ follows from the spectroscopy of hydrogen-like ions. Table~\ref{tab:scales} and Fig.~\ref{fig:scales} collect the scales.

\textit{Isotopes with spin in King plots.}---Isotopes with nuclear spin can nevertheless serve in the King plots that search for new bosons~\cite{Berengut2018,Berengut2020}, which have relied almost exclusively on spinless isotopes~\cite{BerengutOreshkina2025}. Berengut and Oreshkina showed that the second-order hyperfine shift of two isotopes with equal spin is removed by a specific difference weighted by their squared magnetic moments, and took ${}^{111}$Cd and ${}^{113}$Cd, both with $I=1/2$, as an example~\cite{BerengutOreshkina2025}. The scalar part of this shift has a closed form for any multipolarity; it factorizes into an electronic factor times $\mu^2(I+1)/I$ (M1) or a function of $Q_s$ (E2), and its short-distance part has the form of a field shift~\cite{supplemental}. For the $^1S_0\to{}^3P_0$ clock line of cadmium~\cite{Yamaguchi2019}, a CI+MBPT calculation~\cite{AMBiT} gives a second-order shift of $-0.96$~MHz for ${}^{111}$Cd. The specific difference leaves two terms: the differential hyperfine anomaly of the matrix elements involved, which we compute to be $2.04\times10^{-5}$, $14\%$ above that of the $5s$ electron, and a third-order term that scales as $\mu^3$ and leaves $-11$~Hz. Both can be calculated, and about $2$~Hz remains once they are subtracted~\cite{supplemental}. Isotope shifts on narrow lines are now measured with uncertainties from below $0.07$~Hz in Ca$^+$~\cite{Wilzewski2025} to below $16$~Hz in Yb$^+$~\cite{Door2025}.

\textit{Conclusion.}---The gravitational couplings of a nuclear spin to a bound electron are $38$ to $39$ orders of magnitude below the hyperfine interactions that share their tensor and radial structure [Eq.~(\ref{eq:ratios})]. Angular momentum decides which levels carry them (Barrier I) but offers no way to isolate them: within a level they are absorbed into the hyperfine constants (II), centroids cancel them (III), and the $2p$ lines of hydrogen-like ions do not even resolve the hyperfine structure (IV). Barriers I--III use only rotational invariance and the rank of the coupling, so they hold for any coupling of the nuclear spin to the electron, from torsion and Lorentz-violating backgrounds to the exchange of new bosons; what is specific to gravity is only its size~\cite{supplemental}. Ratios of the same size hold in any electronic atom or ion, and they are only $m_\mu/m_e\approx207$ times larger in muonic atoms, which is why laboratory tests of the gravitational coupling of spin use macroscopic sources of gravity and rotation, such as the Earth~\cite{Venema1992,SilenkoTeryaev2007,Heckel2008,Zhang2023}, rather than the nucleus.

\begin{acknowledgments}
Discussions with Dr.~Leonardo F.~Calder\'on and Jos\'e Miguel Mu\~noz are acknowledged with pleasure. This work was supported by the R+D+I efforts from guane Enterprises.
\end{acknowledgments}

%

\clearpage
\onecolumngrid
\setcounter{section}{0}
\setcounter{secnumdepth}{3}   
\renewcommand{\theHsection}{S\arabic{section}}
\renewcommand{\theHequation}{S\arabic{equation}}
\renewcommand{\theHtable}{S\arabic{table}}
\renewcommand{\theHfigure}{S\arabic{figure}}
\begin{center}
{\large\bfseries Supplemental Material for ``Electromagnetic Backgrounds and Angular-Momentum Barriers in Gravitomagnetic Searches with Hydrogen-Like Ions''\par}
\vspace{1.2em}
Leonardo A. Pach\'{o}n\\[3pt]
{\small\itshape guane Enterprises, R+D+I Unit, Medell\'in 050010, Colombia}\\[3pt]
{\small (Dated: October 18, 2026)}
\end{center}
\vspace{1em}
\setcounter{equation}{0}
\setcounter{figure}{0}
\setcounter{table}{0}
\renewcommand{\theequation}{S\arabic{equation}}
\renewcommand{\thefigure}{S\arabic{figure}}
\renewcommand{\thetable}{S\arabic{table}}
\renewcommand{\thesection}{S\arabic{section}}

This Supplemental Material collects the calculations behind the Letter. Section~\ref{sec:dirac} fixes the Dirac--Coulomb conventions and gives the radial expectation values of the $2p_{3/2}$ level of hydrogen-like Mo$^{41+}$. Section~\ref{sec:gravity} estimates the gravitational couplings between the nucleus and the bound electron and compares them with the hyperfine couplings of the same tensor structure. Section~\ref{sec:angular} derives the angular-momentum statements (selection rule, degeneracy, centroid cancellation), which hold for any coupling of the nuclear spin. Section~\ref{sec:hfs} gives the magnetic-dipole and electric-quadrupole hyperfine constants and the second-order hyperfine mixing of the $1s$ and $2p$ levels. Section~\ref{sec:width} gives the radiative widths, Sec.~\ref{sec:data} the nuclear data, and Sec.~\ref{sec:kingodd} how isotopes with spin can enter the King plots used to search for new bosons. Constants are CODATA 2022~\cite{S-CODATA2022}; nuclear data are from the IAEA LiveChart of Nuclides~\cite{S-IAEALiveChart}, which draws on Refs.~\cite{S-Angeli2013,S-StoneQ2021,S-StoneMu2025}.

\section{Dirac--Coulomb states}
\label{sec:dirac}

We use the conventions of Ref.~\cite{S-Johnson2007}. For a point nucleus of charge $Ze$, the bound state $|n\kappa m\rangle$ is
\begin{equation}
    \psi_{n\kappa m}=\frac{1}{r}\begin{pmatrix} P(r)\,\Omega_{\kappa m}\\ iQ(r)\,\Omega_{-\kappa m}\end{pmatrix},\qquad \int_0^\infty(P^2+Q^2)\,dr=1,
    \label{eq:spinor}
\end{equation}
with energy
\begin{equation}
    \frac{E_{n\kappa}}{m_ec^2}=\left[1+\frac{(Z\alpha)^2}{(n-|\kappa|+\gamma)^2}\right]^{-1/2},\qquad \gamma=\sqrt{\kappa^2-(Z\alpha)^2},
    \label{eq:dirac_energy}
\end{equation}
where $\kappa=-1$ ($s_{1/2}$), $+1$ ($p_{1/2}$), $-2$ ($p_{3/2}$). For $Z=42$, $E_{2p_{3/2}}/m_ec^2=0.988188$ and $E_{2p_{1/2}}/m_ec^2=0.987895$, so the $2p$ fine-structure splitting is $\Delta E_{\rm FS}=149.73$~eV; its leading-order value $(Z\alpha)^4m_ec^2/32=140.9$~eV is $6\%$ lower.

The radial functions are known in closed form~\cite{S-Johnson2007}. For a nodeless level ($n=|\kappa|$, such as $1s_{1/2}$ and $2p_{3/2}$) both components are proportional to $r^{\gamma}e^{-\lambda r}$ with $\lambda=Z\alpha\,m_ec/(|\kappa|\hbar)$, so that
\begin{equation}
    \langle r^{-k}\rangle=(2\lambda)^{k}\,\frac{\Gamma(2\gamma+1-k)}{\Gamma(2\gamma+1)} .
    \label{eq:rinv}
\end{equation}
For $2p_{3/2}$ ($\gamma=1.9764$) this gives
\begin{equation}
    \langle r^{-3}\rangle=2.194\times10^{34}\ \text{m}^{-3},\qquad
    \langle r^{-4}\rangle=1.827\times10^{46}\ \text{m}^{-4}.
    \label{eq:rinv_values}
\end{equation}
The first is $1.053$ times the nonrelativistic value $Z^3/(24a_0^3)$. Only $0.57\%$ of $\langle r^{-4}\rangle$ comes from $r<5.5$~fm: for the $2p_{3/2}$ level both expectation values are set by atomic distances, and the finite size of the nucleus is irrelevant at the level of accuracy needed here. At the nuclear surface the $2p_{3/2}$ radial density is small: $(P^2+Q^2)/r^2=5.5\times10^{29}$~m$^{-3}$ at $r=5.5$~fm, against $8.7\times10^{33}$~m$^{-3}$ for $2p_{1/2}$, whose small component has the angular structure of an $s$ wave.

\section{Gravitational couplings of a spinning nucleus}
\label{sec:gravity}

\textit{Multipole structure.}---The Geroch--Hansen multipole moments of the Kerr metric are $M_\ell+iJ_\ell=M(ia)^\ell$, with $a=J/(Mc)$~\cite{S-Hansen1974}. The spin therefore enters as a current (gravitomagnetic) dipole $J_1=Ma$ and as a mass (gravitoelectric) quadrupole $M_2=-Ma^2$; the current quadrupole vanishes, $J_2=0$. For a material body only the first statement is general: its mass quadrupole depends on its structure. In the weak-field limit and at first order in $J$, the field of a rotating body is the Lense--Thirring field, in which a gyroscope precesses with~\cite{S-Mashhoon2003}
\begin{equation}
    \bm{\Omega}_{\rm LT}=\frac{G}{c^2r^3}\left[3(\mathbf{J}\cdot\hat{\mathbf{r}})\hat{\mathbf{r}}-\mathbf{J}\right],
    \label{eq:omega_lt}
\end{equation}
and the $J^2$-dependent part is the Newtonian potential of the quadrupole, $\delta\Phi=-GM_2P_2(\cos\theta)/r^3$.

\textit{Lense--Thirring coupling.}---A slowly moving electron couples to Eq.~(\ref{eq:omega_lt}) through its spin, $H_S=\bm{\Omega}_{\rm LT}\cdot\mathbf{S}$~\cite{S-Mashhoon2003}, and in its orbital motion through the gravitomagnetic vector potential $\mathbf{A}=G\,\mathbf{J}\times\mathbf{r}/(cr^3)$, whose curl gives $\bm{\Omega}_{\rm LT}=\nabla\times\mathbf{A}/c$: Eq.~(\ref{eq:omega_lt}) for $r\neq0$ and, at the origin, the contact term $(8\pi G/3c^2)\,\mathbf{J}\,\delta^3(\mathbf{r})$, the analog of the Fermi contact field. The test-particle Lagrangian of Ref.~\cite{S-Mashhoon2003} couples the velocity to $2\mathbf{A}$, which gives $H_L=(2/c)\,\mathbf{A}\cdot\mathbf{p}=2G\,\mathbf{J}\cdot\mathbf{L}/(c^2r^3)$. A gyrogravitational ratio $\chi\neq1$ rescales $H_S$. With $\mathbf{J}=\hbar\mathbf{I}$, $\mathbf{L}=\hbar\bm{\ell}$ and $\mathbf{S}=\hbar\mathbf{s}$,
\begin{equation}
    H_{\rm LT}=\frac{G\hbar^2}{c^2r^3}\,\mathbf{I}\cdot\mathbf{N}_G,\qquad \mathbf{N}_G=2\bm{\ell}-\mathbf{s}+3(\mathbf{s}\cdot\hat{\mathbf{r}})\hat{\mathbf{r}},
    \label{eq:h_lt}
\end{equation}
while the magnetic-dipole hyperfine interaction of an electron with $\ell\ge1$ (nonrelativistic, with $g_s=2$ and without the contact term) is $(\mu_0/4\pi)\,2\mu_B(\mu/I)\,r^{-3}\,\mathbf{I}\cdot\mathbf{N}_M$ with $\mathbf{N}_M=\bm{\ell}-\mathbf{s}+3(\mathbf{s}\cdot\hat{\mathbf{r}})\hat{\mathbf{r}}$: the two differ only in the weight of the orbital term. Within a level $j$, $\langle\mathbf{N}_M\cdot\mathbf{j}\rangle=f_M\,j(j+1)$ with $f_M=\ell(\ell+1)/[j(j+1)]$, and $\langle\mathbf{N}_G\cdot\mathbf{j}\rangle=f_G\,j(j+1)$ with $f_G=f_M+\langle\bm{\ell}\cdot\mathbf{j}\rangle/[j(j+1)]$. For $2p_{3/2}$, $f_M=8/15$ and $f_G=6/5$; for $2p_{1/2}$, $f_M=8/3$ and $f_G=4$. With the Dirac value of $\langle r^{-3}\rangle$, the Lense--Thirring contribution to the magnetic-dipole constant of $2p_{3/2}$ is
\begin{equation}
    A_{\rm grav}=\frac{G\hbar^2}{c^2}\,\langle r^{-3}\rangle\,f_G=1.36\times10^{-42}\ \text{eV},
    \label{eq:a_grav}
\end{equation}
of which the spin part, the one multiplied by $\chi$, is $(G\hbar^2/c^2)\langle r^{-3}\rangle(f_M-\langle\bm{\ell}\cdot\mathbf{j}\rangle/[j(j+1)])=-1.51\times10^{-43}$~eV. The angular structure is nonrelativistic, so these values carry relative corrections of order $(Z\alpha)^2\approx0.09$. The ratio to the magnetic constant of Table~\ref{tab:hfs} is $\epsilon_1=|A_{\rm grav}/A|=5.6\times10^{-39}$ for ${}^{95}$Mo ($\mu=-0.9097\,\mu_N$). It is the ratio of the prefactors,
\begin{equation}
    \frac{G\hbar^2I/c^2}{(\mu_0/4\pi)\,2\mu_B|\mu|}=2.4\times10^{-39},
    \label{eq:ratio_dip}
\end{equation}
times $f_G/f_M=9/4$ and the factor $1.024$ by which the nonrelativistic $A$, evaluated with the Dirac $\langle r^{-3}\rangle$, exceeds the Dirac value of Table~\ref{tab:hfs}. The angular factor becomes $3/2$ for $2p_{1/2}$. In the $1s$ level only the spin contributes, through the contact term of $\bm{\Omega}_{\rm LT}$, which has the form of the Fermi contact term, so that the ratio is Eq.~(\ref{eq:ratio_dip}) itself: $A_{\rm grav}(1s)=2.4\times10^{-39}\,|A(1s)|\,\chi=8.0\times10^{-41}\chi$~eV. This is a nonrelativistic estimate, and the relativistic factors of the two couplings differ. That of the magnetic constant, $1/[\gamma(2\gamma-1)]=1.16$ for $1s$, comes from the odd operator $\bm{\alpha}\cdot\mathbf{A}$. The coupling of the spin to $\bm{\Omega}_{\rm LT}$ involves the even operator $\bm{\Sigma}$, so its contact term samples the density $(P^2+Q^2)/(4\pi r^2)$ at the nucleus, which for a point nucleus diverges as $r^{2\gamma-2}$. At $r=5.6$~fm, the radius of the uniform sphere with the charge radius of ${}^{95}$Mo, this density is $1.70$ times the nonrelativistic value $Z^3/(\pi a_0^3)$. $A_{\rm grav}(1s)$ may therefore be about $1.5$ times larger, by an amount that depends on the distribution of the nuclear mass current (the gravitational analog of the Bohr--Weisskopf effect); the order of magnitude of the bound below does not change. A renormalization of $\mu$ would change $A$ by the same fraction in every level. The Lense--Thirring term changes it by fractions that differ from level to level by factors of order unity, so its level dependence is a hyperfine anomaly of relative size $\sim10^{-39}$, far below the Bohr--Weisskopf and QED corrections.

\textit{Mass quadrupole.}---The quadrupole enters as the potential energy $m_e\,\delta\Phi$, which has the form of the electric-quadrupole interaction with $e^2Q_s/(4\pi\varepsilon_0)$ replaced by $Gm_eQ_m$, where $Q_m=\int\rho_m(3z^2-r^2)\,d^3r=2M_2$ is the mass quadrupole in the spectroscopic convention. A real nucleus is not a Kerr source: its mass quadrupole is set by its deformation. If mass and charge have the same shape, $Q_m=(M/Z)\,Q_s$, and
\begin{equation}
    \epsilon_2=\frac{B_{\rm grav}}{B}=\frac{Gm_eM}{Ze^2/(4\pi\varepsilon_0)}=9.9\times10^{-40},
    \label{eq:ratio_quad}
\end{equation}
so that $B_{\rm grav}=-2.75\times10^{-44}$~eV for ${}^{95}$Mo and $+3.3\times10^{-43}$~eV for ${}^{97}$Mo. In ${}^{95}$Mo, whose quadrupole moment comes from an odd neutron, the neutrons contribute to $Q_m$ but not to $Q_s$, so this estimate holds only to within a factor of order unity. The spin-induced (Kerr) quadrupole, $Q_m=2M_2=-2J^2/(Mc^2)$, corresponds to a change of the quadrupole moment by $\delta Q=-8\pi\varepsilon_0Gm_eJ^2/(e^2Mc^2)=-2.6\times10^{-44}$~b, i.e., to $-3.3\times10^{-47}$~eV in $B(2p_{3/2})$ of ${}^{95}$Mo; for a quantum spin only the rank-2 part of $I_iI_j$ acts within the level, which reduces this by $(2I-1)/(2I)=4/5$, to $-2.6\times10^{-47}$~eV. Because gravity couples to the energy--momentum of the electron ($\beta mc^2$ and kinetic terms) rather than to its charge, these ratios also carry relative corrections of order $(Z\alpha)^2$, which differ from level to level: for $2p_{3/2}$, the coupling to $\beta mc^2$ weights the radial density with $P^2-Q^2$ instead of $P^2+Q^2$, which multiplies $\langle r^{-3}\rangle$ by $\gamma/|\kappa|=0.988$. The gravitational quadrupole term is a renormalization of $Q_s$ only to this accuracy.

\textit{Terms independent of the nuclear spin.}---The Newtonian energy $GMm_e\langle r^{-1}\rangle$, with the Dirac value $\langle r^{-1}\rangle=\lambda/\gamma$ of Eq.~(\ref{eq:rinv}), is $1.2\times10^{-35}$~eV for $2p_{3/2}$ and $5.0\times10^{-35}$~eV for $1s_{1/2}$, so it shifts the $1s_{1/2}\to2p_{3/2}$ line by $3.8\times10^{-35}$~eV. Between ${}^{95}$Mo and ${}^{97}$Mo, the level energy of $2p_{3/2}$ differs by $2.5\times10^{-37}$~eV and the shift of the line by $8.0\times10^{-37}$~eV. These terms do not depend on the nuclear spin and survive in the centroids (Sec.~\ref{sec:angular}). So do two smaller ones. The geodetic (de Sitter) spin--orbit coupling of the electron in the field of the nucleus, $(3/2)GM\hbar^2\langle r^{-3}\rangle\langle\bm{\ell}\cdot\mathbf{s}\rangle/(m_ec^2)$, adds $+1.5\times10^{-37}$~eV to $2p_{3/2}$ and vanishes for $1s$. The post-Newtonian corrections to the monopole coupling, which the Lagrangian of Ref.~\cite{S-Mashhoon2003} gives as $m\gamma(1+v^2/c^2)\Phi$, have relative size $\sim(3/2)\langle v^2/c^2\rangle=(3/2)(Z\alpha/n)^2$: about $4\times10^{-37}$~eV for $2p_{3/2}$ and $7\times10^{-36}$~eV for $1s$.

\textit{Spin and the quadrupole.}---A static coupling of the electron spin to a field that depends on $I^2$, such as the coupling $\mathbf{S}\cdot\mathbf{B}_g$, with $B_g\propto I^2$, couples a rank-2 nuclear tensor, which is even under time reversal, to an electronic operator $[C^{(2)}\otimes\mathbf{s}]^{(2)}$, which is odd. Its diagonal matrix elements vanish; for $2p_{3/2}$ this is the vanishing of the $9j$ symbol $\{1\,1\,2;\tfrac12\,\tfrac12\,1;\tfrac32\,\tfrac32\,2\}$ (Sec.~\ref{sec:angular}). The spin couples to the mass quadrupole only through velocity-dependent (spin--orbit) terms, which are further suppressed by powers of $Z\alpha$.

\textit{Bound on $\chi$.}---A measured hyperfine constant constrains $\chi-1$ only through an independent prediction of its magnetic part, and for a single level the nuclear moment alone limits that prediction. For ${}^{95}$Mo, $\delta\mu/\mu=3.3\times10^{-4}$~\cite{S-IAEALiveChart} makes $A(1s)$ uncertain by $|A(1s)|\,\delta\mu/\mu=1.1\times10^{-5}$~eV and $A(2p_{3/2})$ by $8.0\times10^{-8}$~eV. With the spin parts $8.0\times10^{-41}\chi$~eV and $-1.51\times10^{-43}\chi$~eV, this gives $|\chi-1|\lesssim1.4\times10^{35}$ from $1s$, which is $\delta\mu/\mu$ divided by Eq.~(\ref{eq:ratio_dip}), and $5\times10^{35}$ from $2p_{3/2}$, whose constant cannot be measured in the $1s\to2p_{3/2}$ line (Sec.~\ref{sec:width}). At this level the bound applies equally to $\chi$ and to $\chi-1$. It is a best case: it takes the tabulated $\delta\mu/\mu$ at face value and ignores the Bohr--Weisskopf and QED corrections, which would have to be predicted to $10^{-39}$ of $A$. The tabulated value itself has moved: the recommended moment of ${}^{95}$Mo was $-0.9132(3)\,\mu_N$ in 2019~\cite{S-StoneMu2019} and is $-0.9097(3)\,\mu_N$ since its re-evaluation with a calculated diamagnetic shielding~\cite{S-StoneMu2025,S-AntusekRepisky2020}. The change, $3.8\times10^{-3}$, is twelve times the quoted uncertainty; an uncertainty of that size would weaken the $1s$ bound to $|\chi-1|\lesssim2\times10^{36}$. Ratios of constants of different levels do not depend on $\mu$, but they require the same corrections, for both levels, at the same level of accuracy.

Laboratory tests do not probe the field of a spinning source, which has not been measured for intrinsic spins~\cite{S-Fadeev2021}, but the coupling of spin to the rotation of the laboratory, which the equivalence principle relates to $\chi$~\cite{S-SilenkoTeryaev2007}. For electron spins, the gyrocompass torque of the Earth's rotation on a torsion pendulum with $\approx10^{23}$ polarized electrons, measured to $2.2\%$, agrees with the spin content estimated from its magnetization, which is known to $18\%$ ($1\sigma$)~\cite{S-Heckel2008,S-Ni2010}. The torque acts on the total angular momentum of the pendulum, $\mathbf{J}=-\mathbf{S}$, whose orbital part is $\mathbf{L}=-2\mathbf{S}$, so it tests the coupling of spin given the standard coupling of orbital angular momentum; Ref.~\cite{S-Heckel2008} uses the effect as a calibration, and Ref.~\cite{S-Ni2010} reads it as a test. With nuclear spins, a ${}^{129}$Xe/${}^{131}$Xe comagnetometer has measured the Earth's rotation rate to $3\times10^{-4}$~\cite{S-Zhang2023}.

\textit{Rotation coupling of the ${}^{129}$Xe/${}^{131}$Xe pair.}---If each species precesses at $\omega_i=\gamma_iB+(1+\zeta_i)\Omega_E\cos\theta$, the convention of Eq.~(3) of Ref.~\cite{S-Zhang2023}, where $\theta$ is the angle between $\mathbf{B}$ and the Earth's axis and $\zeta_i$ is an anomaly in the coupling of its spin to the rotation of the laboratory, the rotation rate extracted from the ratio of the two frequencies with $\zeta_i=0$ is $\Omega_{E,m}=\Omega_E[1+(\zeta_{129}-\rho\zeta_{131})/(1-\rho)]$, where $\rho=\gamma_{129}/\gamma_{131}=-3.37337(2)$~\cite{S-Zhang2023}; the sign of the rotation term does not affect this combination. The measured Earth rotation rate, $\Omega_{E,m}/2\pi=11605.7\pm2.5\,({\rm stat})\pm0.6\,({\rm sys})$~nHz, obtained after the correction for the orientation of the field (Ref.~\cite{S-Zhang2023} denotes by $\Omega_m$ the rate resolved before it), against $11605.761$~nHz~\cite{S-Zhang2023}, gives $(\zeta_{129}-\rho\zeta_{131})/(1-\rho)=(-0.05\pm2.2)\times10^{-4}$, so that $|\zeta_{129}-\rho\zeta_{131}|/(1-\rho)<4.5\times10^{-4}$ at $2\sigma$. With the same normalization, the Hg bound becomes $0.042/1.369=3.1\times10^{-2}$ at $95\%$, about 70 times weaker. Two caveats apply. In that experiment the rotation term ($\propto\cos\theta$) and a coupling of spin to gravity, $\propto\bm{\sigma}\cdot\mathbf{g}$ ($\propto\cos\phi$, with $\phi$ the angle between $\mathbf{B}$ and $\mathbf{g}$), reverse together with $\mathbf{B}$, so a single geometry bounds only one combination: reading it as a bound on $\zeta_i$ assumes that there is no spin--gravity coupling, just as the bound on the latter in Ref.~\cite{S-Zhang2023} assumes the standard rotation coupling. And reading $\zeta_i$ as an anomalous gravitomagnetic moment rests on the equivalence-principle argument of Ref.~\cite{S-SilenkoTeryaev2007}, whereas Ref.~\cite{S-Obukhov2014} argues that such a moment cannot be introduced covariantly; the bound is best read as a test of the coupling of spin to rotation.

\section{Angular momentum algebra}
\label{sec:angular}

Every coupling between the nucleus and the electron considered here has the form $H_k=\mathbf{t}^{(k)}\cdot\mathbf{T}^{(k)}$, with $\mathbf{T}^{(k)}$ a rank-$k$ nuclear tensor and $\mathbf{t}^{(k)}$ an electronic one ($k=1$: M1 and Lense--Thirring; $k=2$: E2 and the gravitational quadrupole). In the hyperfine basis~\cite{S-Varshalovich1988},
\begin{equation}
    \langle(jI)F|H_k|(j'I)F\rangle=(-1)^{j'+I+F}
    \begin{Bmatrix} j & I & F\\ I & j' & k\end{Bmatrix}
    \langle j\|t^{(k)}\|j'\rangle\langle I\|T^{(k)}\|I\rangle .
    \label{eq:racah}
\end{equation}

\textit{Selection rule.}---The diagonal element ($j'=j$) vanishes unless $k\le2j$ and $k\le2I$. A rank-2 coupling therefore needs $j\ge1$, i.e., $j\ge3/2$ for a single electron; the $j=1/2$ levels ($1s_{1/2}$, $2s_{1/2}$, $2p_{1/2}$) see only the rank-0 and rank-1 parts.

\textit{Time reversal.}---The nuclear quadrupole tensor is even under time reversal, so a hermitian coupling to it needs an even electronic operator of rank 2. An operator built from the electron spin and a static spatial tensor, $[C^{(2)}\otimes\mathbf{s}]^{(2)}$, is odd, and its diagonal matrix elements vanish: in the $2p_{3/2}$ level they are proportional to $\{1\,1\,2;\tfrac12\,\tfrac12\,1;\tfrac32\,\tfrac32\,2\}=0$. The velocity-dependent $[\bm{\ell}\otimes\mathbf{s}]^{(2)}$ is even and survives, with $\{1\,1\,1;\tfrac12\,\tfrac12\,1;\tfrac32\,\tfrac32\,2\}=\sqrt6/36$.

\textit{Degeneracy.}---Within a level of fixed $j$ and $I$, the $F$ dependence of Eq.~(\ref{eq:racah}) is fixed by the $6j$ symbol and is the same for every coupling of the same rank. For $k=1$ it is the interval rule $\propto K=F(F+1)-I(I+1)-j(j+1)$ of the magnetic-dipole constant $A$; for $k=2$ it is the pattern of the electric-quadrupole constant $B$,
\begin{equation}
    \frac{E_Q(F)}{B}=\frac{\tfrac34K(K+1)-I(I+1)j(j+1)}{2I(2I-1)j(2j-1)}.
    \label{eq:casimir}
\end{equation}
A gravitational rank-1 (rank-2) term thus only adds to $A$ ($B$); it cannot be separated from the electromagnetic term within one level.

\textit{Centroid.}---Summing Eq.~(\ref{eq:racah}) with weights $2F+1$ gives the trace of $\mathbf{t}^{(k)}\cdot\mathbf{T}^{(k)}$ over the $(2j+1)(2I+1)$ states of the level, $\sum_q(-1)^q\,\mathrm{Tr}\,t^{(k)}_q\,\mathrm{Tr}\,T^{(k)}_{-q}$, which vanishes for $k\ge1$ because every component of an irreducible tensor of rank $k\ge1$ is traceless. The $(2F+1)$-weighted centroid of a hyperfine multiplet is therefore unshifted, at first order, by M1, by E2, and by any gravitational coupling of rank 1 or 2. For $I=5/2$, $j=3/2$ the pattern~(\ref{eq:casimir}) is $E_Q/B=\tfrac{7}{10},-\tfrac{1}{10},-\tfrac{11}{20},\tfrac14$ for $F=1,\dots,4$, with centroid $0$. The same holds for the centroid of any line $J\to J'$ of multipolarity $L$ weighted by its line strengths $\propto(2F+1)(2F'+1)\{J\,F\,I;F'\,J'\,L\}^2$, for every rank $k\ge1$: since $\sum_F(2F+1)\{J\,F\,I;F'\,J'\,L\}^2=1/(2J'+1)$ for each $F'$, the line centroid is the difference of the $(2F+1)$-weighted centroids of the two levels.

\textit{Couplings beyond general relativity.}---The selection rule, the degeneracy and the centroid cancellation use only rotational invariance and the rank $k$ of the coupling. They hold for any coupling of the nuclear spin to the bound electron, whatever its origin: torsion, Lorentz-violating backgrounds, or the exchange of new bosons. Within a level, a coupling of rank $k$ only shifts the hyperfine constant of that rank. If it also has the radial dependence and the orbital-to-spin weighting of the hyperfine interaction, it shifts the constants of all levels in the same proportion and cannot be told apart from a change of $\mu$ or $Q_s$. A coupling with another radial dependence, of contact or finite range, or with another weighting changes the ratios between levels, and only such ratios can reveal it, as in the specific difference between the hyperfine splittings of lithium-like and hydrogen-like ions of Ref.~\cite{S-Shabaev2001}; it then competes with the uncertainty of the magnetic parts, as the Lense--Thirring term does (Sec.~\ref{sec:gravity}). Gravity differs from these couplings only in its size.

\section{Hyperfine structure of the $1s$ and $2p$ levels}
\label{sec:hfs}

The magnetic-dipole and electric-quadrupole operators have reduced matrix elements~\cite{S-Johnson2007}
\begin{align}
    \langle\kappa_a\|t^{(1)}\|\kappa_b\rangle&=-(\kappa_a+\kappa_b)\,\langle-\kappa_a\|C_1\|\kappa_b\rangle\int_0^\infty\frac{P_aQ_b+Q_aP_b}{r^2}\,dr,\label{eq:t1}\\
    \langle\kappa_a\|t^{(2)}\|\kappa_b\rangle&=-\langle\kappa_a\|C_2\|\kappa_b\rangle\int_0^\infty\frac{P_aP_b+Q_aQ_b}{r^3}\,dr,\label{eq:t2}
\end{align}
with the prefactors $(\mu_0/4\pi)ec$ and $e^2/(4\pi\varepsilon_0)$ and the nuclear reduced matrix elements $\langle I\|\mu\|I\rangle=\mu\sqrt{(2I+1)(I+1)/I}$ and $\langle I\|Q\|I\rangle=\tfrac12Q_s\sqrt{(2I+1)(I+1)(2I+3)/[I(2I-1)]}$. For a single level they reduce to
\begin{equation}
    A_\kappa=\frac{\mu_0}{4\pi}\frac{ec\,\mu}{I}\,\frac{\kappa}{j(j+1)}\int_0^\infty\frac{2PQ}{r^2}\,dr,\qquad
    B=\frac{e^2Q_s}{4\pi\varepsilon_0}\,\frac{2j-1}{2j+2}\,\langle r^{-3}\rangle .
    \label{eq:AB}
\end{equation}
We evaluate the radial integrals with the analytic point-nucleus functions and check that Eq.~(\ref{eq:racah}) reproduces $AK/2$ and Eq.~(\ref{eq:casimir}) to $10^{-15}$. Table~\ref{tab:hfs} lists the results. The $2p_{3/2}$ levels with $F=1$ and $F=4$ do not mix with $2p_{1/2}$, whose hyperfine levels have $F=2,3$. The second-order shift of the $2p_{3/2}$ level $F$ is $|\langle2p_{1/2}F|H_1+H_2|2p_{3/2}F\rangle|^2/\Delta E_{\rm FS}$, and its $(2F+1)$-weighted centroid is the scalar hyperfine shift that survives in the centroid of the $1s\to2p_{3/2}$ line. It is dominated by M1--M1 mixing. The lower level of the line, $1s_{1/2}$, has $A(1s)=-3.30\times10^{-2}$~eV for ${}^{95}$Mo (splitting $0.099$~eV); E2 does not act on it ($k=2>2j$), but M1 mixes it with $2s_{1/2}$, which lowers its centroid by $1.71\times10^{-8}$~eV (${}^{95}$Mo) and $1.78\times10^{-8}$~eV (${}^{97}$Mo). Higher $ns$ levels and the positive-energy continuum push the $1s$ level in the same direction, so within the no-pair approximation these values are lower bounds on the magnitude; $ns$ levels up to $n=6$ raise them to $2.44\times10^{-8}$~eV and $2.55\times10^{-8}$~eV (Table~\ref{tab:hfs}). These bound-state sums are not representative of the total, which the continuum dominates. In the nonrelativistic limit the second-order contact interaction is proportional to $\sum_k|\psi_k(0)|^2/(E_k-E_{1s})$, whose continuum part diverges linearly in the momentum $\hbar k$ of the continuum states. With Coulomb continuum functions, normalized so that the density of $|\psi(0)|^2$ per unit energy is continuous at the threshold, the bound $ns$ levels with $n\ge2$ give $1.5$ times the $2s$ term, the continuum with $k<Z/a_0$ gives $2.1$ times, and the continuum up to $k=m_ec/\hbar$ gives $7.7$ times. Above $m_ec/\hbar$ the nonrelativistic estimate fails: the sum is cut off by the distribution of the nuclear magnetization, and the negative-energy states, absent in the no-pair approximation, contribute with the opposite sign. The second-order shift of the $1s$ centroid is therefore $\sim10^{-7}$~eV or more (about $1.6\times10^{-7}$~eV for ${}^{95}$Mo with the continuum up to $k=m_ec/\hbar$) and depends on the structure of the nucleus; a reliable value requires a Dirac--Coulomb Green's-function calculation with a finite nuclear magnetization. The second-order hyperfine shift of the centroid of the $1s\to2p_{3/2}$ line is therefore dominated by the $1s$ level. Finite-nuclear-size (Bohr--Weisskopf) and QED corrections are not included.

\begin{table}[t]
\caption{\label{tab:hfs} Hyperfine structure of the $1s$ and $2p$ levels of hydrogen-like Mo$^{41+}$ (point nucleus, energies in eV). $\delta B$ is the uncertainty of $B$ from that of $Q_s$. The rows M1--M1, E2--E2 and total are the $(2F+1)$-weighted centroids of the second-order shifts of $2p_{3/2}$; the last two rows are those of $1s_{1/2}$ from mixing with $2s$ and with $ns$ up to $n=6$.}
\begin{ruledtabular}
\begin{tabular}{lcc}
 & ${}^{95}$Mo & ${}^{97}$Mo \\
\hline
$A(2p_{1/2})$ & $-1.437\times10^{-3}$ & $-1.467\times10^{-3}$ \\
$A(2p_{3/2})$ & $-2.430\times10^{-4}$ & $-2.481\times10^{-4}$ \\
$B(2p_{3/2})$ & $-2.780\times10^{-5}$ & $+3.222\times10^{-4}$ \\
$\delta B$ & $1.3\times10^{-6}$ & $1.6\times10^{-5}$ \\
$A(1s_{1/2})$ & $-3.30\times10^{-2}$ & $-3.37\times10^{-2}$ \\
M1, $F=2$ & $-2.92\times10^{-4}$ & $-2.98\times10^{-4}$ \\
M1, $F=3$ & $-3.49\times10^{-4}$ & $-3.56\times10^{-4}$ \\
E2, $F=2$ & $+2.22\times10^{-5}$ & $-2.57\times10^{-4}$ \\
E2, $F=3$ & $-1.33\times10^{-5}$ & $+1.54\times10^{-4}$ \\
M1--M1 & $3.6\times10^{-10}$ & $3.7\times10^{-10}$ \\
E2--E2 & $1.0\times10^{-12}$ & $1.4\times10^{-10}$ \\
total & $3.6\times10^{-10}$ & $5.1\times10^{-10}$ \\
$1s$ centroid ($2s$ only) & $-1.71\times10^{-8}$ & $-1.78\times10^{-8}$ \\
$1s$ centroid ($ns$, $n\le6$) & $-2.44\times10^{-8}$ & $-2.55\times10^{-8}$ \\
\end{tabular}
\end{ruledtabular}
\end{table}

\textit{Centroid of an unresolved line.}---When the hyperfine structure is not resolved (Sec.~\ref{sec:width}), a measurement determines the intensity-weighted centroid of the line. With line-strength weights and statistical populations it is unshifted at first order (Sec.~\ref{sec:angular}). At second order, besides the level shifts, which move it by $+1.7\times10^{-8}$~eV for ${}^{95}$Mo with $2s$ alone, the hyperfine mixing changes the E1 amplitudes of the components. The $2p_{1/2}$ admixture in $2p_{3/2}$ interferes with the main amplitude through $\langle1s\|rC_1\|2p_{1/2}\rangle$ and moves the centroid by $+1.0\times10^{-7}$~eV. The $2s$ admixture in $1s$ interferes through $\langle2s\|rC_1\|2p_{3/2}\rangle$, which is $4.1$ times larger in magnitude than $\langle1s\|rC_1\|2p_{3/2}\rangle$, and moves it by $-3.8\times10^{-7}$~eV. Together they give $-2.6\times10^{-7}$~eV for ${}^{95}$Mo and $-2.7\times10^{-7}$~eV for ${}^{97}$Mo. Higher bound levels change these values by about a quarter: including $ns$ up to $n=6$ in $1s$ and $np$ up to $n=4$ in $2p_{3/2}$ gives $-2.0\times10^{-7}$~eV for ${}^{95}$Mo. The continuum, which dominates the second-order shift of the $1s$ level, enters these admixtures as well and was not evaluated, so the magnitude may be several times larger. Moreover, a measured centroid weights the components by their intensities, which carry a factor $\omega^3$ in emission ($\omega$ in absorption) and the populations of the hyperfine sublevels. The factor $\omega^3$ alone shifts the centroid by $+3.8\times10^{-7}$~eV ($\omega$: $+1.3\times10^{-7}$~eV), with the opposite sign to the mixing terms. Non-statistical populations shift it already at first order: an absorption line from the $F=3$ ground sublevel of $1s$ alone is shifted by $+4\times10^{-2}$~eV. These numbers therefore only give a lower scale, $\gtrsim10^{-7}$~eV, for the second-order hyperfine effects on the centroid of the line.

The rows ``M1'' and ``E2'' are the off-diagonal elements $\langle2p_{1/2}F|H_k|2p_{3/2}F\rangle$. The M1--E2 cross term shifts individual levels (for ${}^{97}$Mo it enhances $F=2$ and reduces $F=3$), but it is the product of a rank-1 and a rank-2 nuclear operator, has no scalar part, and drops out of the centroid: the total is the sum of the M1--M1 and E2--E2 centroids.

\section{Transition energies and radiative widths}
\label{sec:width}

From Eq.~(\ref{eq:dirac_energy}), the $1s_{1/2}\to2p_{3/2}$ and $1s_{1/2}\to2p_{1/2}$ transition energies of Mo$^{41+}$ are $18.556$ and $18.407$~keV (the point-nucleus value for $1s_{1/2}\to2s_{1/2}$ coincides with the latter), and $1s_{1/2}\to3d_{5/2}$ is at $21.918$~keV. The NIST Atomic Spectra Database~\cite{S-NISTASD} gives $18.537$ and $18.387$~keV for the $1s\to2p$ lines; the $20$~eV difference comes from the QED, nuclear-size, and recoil corrections that Eq.~(\ref{eq:dirac_energy}) omits. Only the $1s\to2p$ lines are electric-dipole transitions: $1s\to2s$ ($\Delta\ell=0$) and $1s\to3d$ ($\Delta\ell=2$) do not change parity.

The E1 rate of $2p_j\to1s_{1/2}$ in the long-wavelength limit is
\begin{equation}
    A_{\rm E1}=\frac{\omega^3e^2}{3\pi\varepsilon_0\hbar c^3}\,\frac{|\langle1s\|r\,C_1\|2p_j\rangle|^2}{2j+1},
    \label{eq:e1}
\end{equation}
with the radial integral $\int(P_{1s}P_{2p}+Q_{1s}Q_{2p})\,r\,dr$. With the Dirac functions, $A_{\rm E1}(2p_{3/2})=1.99\times10^{15}$~s$^{-1}$ ($\Gamma=\hbar A_{\rm E1}=1.31$~eV) and $A_{\rm E1}(2p_{1/2})=1.95\times10^{15}$~s$^{-1}$ ($\Gamma=1.28$~eV); the nonrelativistic $Z^4$ scaling of the hydrogen Lyman-$\alpha$ rate gives $\Gamma=1.28$~eV. The NIST database~\cite{S-NISTASD} lists $A_{\rm E1}(2p_{3/2})=1.9016\times10^{15}$~s$^{-1}$ and $A_{\rm E1}(2p_{1/2})=1.9668\times10^{15}$~s$^{-1}$, with an accuracy of $1\%$ or better, i.e., $\tau(2p_{3/2})=0.53$~fs and $\Gamma(2p_{3/2})=1.25$~eV. The rates of Eq.~(\ref{eq:e1}) are $4.6\%$ higher than the NIST value for $2p_{3/2}$ and $1.0\%$ lower for $2p_{1/2}$, mostly because of the long-wavelength approximation; $0.3\%$ comes from the Dirac transition energy in $\omega^3$. These widths exceed the $2p$ hyperfine splittings of Table~\ref{tab:hfs} (at most a few $10^{-3}$~eV) by a factor of several hundred.

\section{Nuclear data}
\label{sec:data}

Table~\ref{tab:data} lists the ground-state data used here, as given by the IAEA LiveChart of Nuclides~\cite{S-IAEALiveChart} (extracted on September 24, 2026): charge radii from Ref.~\cite{S-Angeli2013}, magnetic moments from Ref.~\cite{S-StoneMu2025}, which uses the calculated diamagnetic shielding of Ref.~\cite{S-AntusekRepisky2020}, and quadrupole moments from Ref.~\cite{S-StoneQ2021}. The quadrupole moment of ${}^{91}$Mo ($I^\pi=9/2^+$, $t_{1/2}=15.49$~min) is not tabulated; its charge radius and magnetic moment are.

\begin{table}[t]
\caption{\label{tab:data} Ground-state data of the molybdenum isotopes~\cite{S-IAEALiveChart}.}
\begin{ruledtabular}
\begin{tabular}{lcccc}
$A$ & $I^\pi$ & $r_{\rm ch}$ (fm) & $\mu$ ($\mu_N$) & $Q_s$ (b) \\
\hline
91 & $9/2^+$ & 4.3182(12) & $-0.927(3)$ & --- \\
92 & $0^+$ & 4.3151(12) & --- & --- \\
94 & $0^+$ & 4.3529(13) & --- & --- \\
95 & $5/2^+$ & 4.3628(18) & $-0.9097(3)$ & $-0.022(1)$ \\
96 & $0^+$ & 4.3847(15) & --- & --- \\
97 & $5/2^+$ & 4.3880(15) & $-0.9287(4)$ & $+0.255(13)$ \\
98 & $0^+$ & 4.4091(18) & --- & --- \\
100 & $0^+$ & 4.4468(25) & --- & --- \\
\end{tabular}
\end{ruledtabular}
\end{table}

\section{Isotopes with spin in King plots}
\label{sec:kingodd}

The analysis of Sec.~\ref{sec:angular} also shows how isotopes with nuclear spin can enter the King plots used to search for new bosons~\cite{S-Berengut2018,S-Berengut2020}, which have relied almost exclusively on spinless isotopes because of their hyperfine structure~\cite{S-BerengutOreshkina2025}; the King plots of strontium with ${}^{87}$Sr~\cite{S-Miyake2019} and of cadmium with ${}^{111}$Cd and ${}^{113}$Cd~\cite{S-Hofsass2023}, through the centers of gravity of their hyperfine multiplets, are exceptions. At first order the line centroid is unshifted (Sec.~\ref{sec:angular}); what remains is of second order.

\textit{Second-order shift of a centroid.}---Mixing of a level $J$ with a level $J'$ through the hyperfine operators $\mathbf{t}^{(k)}\cdot\mathbf{T}^{(k)}$ shifts its $(2F+1)$-weighted centroid by
\begin{equation}
    \delta_c=\sum_k\frac{\left|\langle J\|t^{(k)}\|J'\rangle\langle I\|T^{(k)}\|I\rangle\right|^2}{(2k+1)(2J+1)(2I+1)(E_J-E_{J'})},
    \label{eq:scalar2}
\end{equation}
which follows from Eq.~(\ref{eq:racah}) and the orthogonality of the $6j$ symbols; for the same reason the M1--E2 cross term has no scalar part. With the nuclear reduced matrix elements of Sec.~\ref{sec:hfs}, the isotope dependence is
\begin{equation}
    \delta_c(A)=G^{(1)}\,\frac{\mu_A^2(I_A+1)}{I_A}+G^{(2)}\,\frac{Q_A^2(I_A+1)(2I_A+3)}{I_A(2I_A-1)},
    \label{eq:isoscaling}
\end{equation}
where the electronic factors $G^{(k)}$ do not depend on the isotope up to the hyperfine anomaly of the off-diagonal matrix elements. For two isotopes with $I=1/2$ only the first term is present, and the specific difference $\delta\nu^{A'}-[s(A')/s(A'')]\,\delta\nu^{A''}$, with $s=\mu^2(I+1)/I$, removes it. For equal spins this is the specific difference weighted by $(\mu_{A'}/\mu_{A''})^2$ proposed by Berengut and Oreshkina, who took ${}^{111}$Cd and ${}^{113}$Cd as an example and noted that two isotopes with spin give one point in a King plot~\cite{S-BerengutOreshkina2025}. What remains is $2\Delta\,|\delta_c|$, with $\Delta$ the differential hyperfine anomaly of the off-diagonal matrix elements, and the terms of third order in the hyperfine interaction, which scale with a higher power of $\mu$. Each odd isotope adds one point to a generalized King plot~\cite{S-Berengut2020} and each nuclear vector in Eq.~(\ref{eq:isoscaling}) one parameter, so tin, with three stable isotopes of spin $1/2$, gains two points, and cadmium and tellurium one, provided the anomaly and the third-order terms are calculated; left as unknowns, each adds a nuclear vector, and cadmium gains none.

\textit{Short-distance part.}---For $s$ levels the sum over intermediate states is dominated by the continuum (Sec.~\ref{sec:hfs}). Its high-energy part has energy denominators close to the energy of the intermediate state, the same for every level, and is the electron density at the nucleus times a nuclear constant, the structure of the field shift. It is therefore absorbed in the field-shift parameter of each odd isotope and does not break King linearity; only the low-energy remainder has to be calculated. A Dirac--Coulomb calculation of the second-order contact term for a nucleus magnetized uniformly within a radius $R_M$ confirms this. For $1s$, $2s$ and $3s$ at $Z=10$, $42$ and $70$, and for $2p_{1/2}$ and $3p_{1/2}$ at $Z=70$, the change of the shift between $R_M=2$ and $16$~fm, divided by the field-shift factor of the level, is the same to within $0.6\%$. At $Z=42$ the $p_{1/2}$ levels deviate by $3\%$, and at $Z=10$ by $24\%$: at lower $Z$ the $p_{1/2}$ density at the nucleus is suppressed by $(Z\alpha)^2$, and finite-size corrections of another form become relatively larger. The test is for one electron in a Coulomb field and does not include core polarization.

\textit{The cadmium clock.}---Cadmium has two stable isotopes with $I=1/2$, ${}^{111}$Cd and ${}^{113}$Cd, and a $^1S_0\to{}^3P_0$ clock transition at $332$~nm~\cite{S-Yamaguchi2019}. In the odd isotopes hyperfine mixing makes it weakly allowed; in the even isotopes it has to be induced by an external field, as for strontium~\cite{S-Miyake2019}, and its isotope shifts have not been measured. With $I=1/2$ only M1 acts, and $5s5p\,{}^3P_0$ mixes with ${}^3P_1$, $542.1$~cm$^{-1}$ above~\cite{S-NISTASDlevels}, and with ${}^1P_1$. We computed the hyperfine matrix elements with the CI+MBPT code AMBiT~\cite{S-AMBiT}, with random-phase-approximation corrections and two valence electrons above a Cd$^{2+}$ core, and normalized them to the measured $A({}^3P_1)$, from the double-resonance intervals quoted in Ref.~\cite{S-Hofsass2023}. The calculation reproduces the measured ratio $A({}^3P_2)/A({}^3P_1)$~\cite{S-Faust1960,S-Hofsass2023} to better than $1\%$, and Eq.~(\ref{eq:scalar2}) gives $\delta_c({}^3P_0)=-0.97$~MHz for ${}^{111}$Cd and $-1.06$~MHz for ${}^{113}$Cd, stable to $1\%$ under an increase of the basis and to $0.2\%$ under a $26\%$ increase of the nuclear charge radius. The ground level shifts by $-6.7$~kHz through the even $J=1$ levels, so that the line shifts by $-0.96$~MHz. The same procedure gives $-350$~kHz for ${}^3P_0$ of ${}^{87}$Sr, normalized to $A({}^3P_1)$ of Ref.~\cite{S-Miyake2019}, against $-322$~kHz in Ref.~\cite{S-BerengutOreshkina2025}; we take the difference, $9\%$, as the accuracy of $\delta_c$. Subtracting a shift of $1$~MHz at the Hz level would require a relative accuracy of $10^{-6}$; the specific difference removes it up to the following terms.

\textit{Hyperfine anomaly.}---For the $5s$ electron, the ground-state splittings of ${}^{111}$Cd$^+$ and ${}^{113}$Cd$^+$~\cite{S-Zhang2012,S-Miao2021} and the ratio of nuclear $g$ factors quoted in Ref.~\cite{S-Hofsass2023} give $\Delta_{5s}=1.785(19)\times10^{-5}$. To transfer it to the matrix elements of $5s5p$, we repeated the calculation for a nucleus magnetized uniformly within a radius $R_M$, changed by $\pm1$~fm, and for the $5s$ level of Cd$^+$ with the same core and basis. The sensitivity of each matrix element to $R_M$, divided by that of $A(5s)$, gives its anomaly in units of $\Delta_{5s}$. For $\delta_c({}^3P_0)$ the ratio is $1.14$, so $\Delta=2.04\times10^{-5}$, stable to $0.3\%$ under changes of the step in $R_M$, the basis, the treatment of correlation and the nuclear charge radius. The ratio exceeds one because in the dominant element, $\langle(5s5p_{1/2})_0\|T\|(5s5p_{1/2})_1\rangle$ in $jj$ coupling, the $5s$ and $5p_{1/2}$ contributions enter with opposite signs, and the $5p_{1/2}$ contribution hardly depends on $R_M$. The same calculation predicts the anomalies of the levels. For ${}^3P_2$ it gives $1.69\times10^{-5}$, against $1.69(6)\times10^{-5}$ from the intervals of Ref.~\cite{S-Faust1960} once the computed second-order shifts are removed ($1.58\times10^{-5}$ before). For ${}^3P_1$ it gives $1.56\times10^{-5}$, against $0.79(45)\times10^{-5}$ ($-0.09\times10^{-5}$ before) from the double-resonance intervals, $1.7\sigma$ apart; we have not consulted the original measurements, and if they already removed the second-order shifts the difference would be $3.7\sigma$. With $|\delta_c|=0.97$~MHz, the residual $2\Delta|\delta_c|$ is $40$~Hz if the anomaly is ignored, $5.0$~Hz if $\Delta_{5s}$ is used, and $1.4$~Hz if the computed $\Delta$ is used, whose uncertainty, $3.5\%$, is set by the test with ${}^3P_2$ and by $\Delta_{5s}$. The second-order shift of the ground level has the anomaly of the $5s$ electron to $0.1\%$ and leaves $0.2$~Hz if the anomaly is ignored.

\textit{Third order.}---With $I=1/2$ the level ${}^3P_0$ has only $F=1/2$ and mixes only with the $F=1/2$ sublevels of the $J=1$ levels, which the hyperfine interaction displaces at first order by $-A$; for ${}^3P_1$ that is $+4.12$~GHz. The shift of ${}^3P_0$ therefore has a third-order part, $\sum_n|V_{0n}|^2(-A_n)/(E_0-E_n)^2=+241$~Hz for ${}^{111}$Cd, which scales as $\mu^3$ and is not removed by the specific difference: it leaves $\delta^{(3)}({}^{111}\mathrm{Cd})[1-g({}^{113}\mathrm{Cd})/g({}^{111}\mathrm{Cd})]=-11.1$~Hz. An exact diagonalization of the $F=1/2$ block gives between $-11.8$ and $-10.5$~Hz, depending on the relative sign of the matrix elements between the $J=1$ levels, which our calculation does not fix. The term can therefore be subtracted with an uncertainty of $1.2$~Hz, from that sign and from the accuracy of $\delta_c$. Once the anomaly and the third-order term are subtracted, the remainders add in quadrature to $1.8$~Hz.

\textit{A counterexample: ${}^{171}$Yb$^+$.}---For ${}^{171}$Yb$^+$ ($I=1/2$), whose quadrupole and octupole transitions from $6s\,{}^2S_{1/2}$ are measured in the even isotopes to below $6$ and $16$~Hz~\cite{S-Door2025}, the ground level itself has a second-order shift. A Dirac--Fock calculation with AMBiT, normalized to the measured $A(6s)$~\cite{S-Fisk1997}, puts its low-energy part, from the bound $ns$ levels and the continuum up to the binding energy of $6s$, at $-18$~kHz. The value depends on how the basis discretizes the continuum: the bound levels give $-11$~kHz, and the next pseudostate of the basis would bring it to $-27$~kHz. It would have to be computed to $10^{-3}$.

\makeatletter
\let\gsqS@label\label
\def\label#1{\def\@tempa{#1}\def\@tempb{LastBibItem}\ifx\@tempa\@tempb\gsqS@label{LastBibItemS}\else\gsqS@label{#1}\fi}
\def\NAT@bibsetnum#1{\setlength{\topsep}{\z@}\NATx@bibsetnum{\ref{LastBibItemS}}}
\makeatother

\end{document}